\documentclass[english,twocolumn,showpacs,preprintnumbers,amsmath,amssymb,aps,prl,floatfix,superscriptaddress]{revtex4}
\usepackage[T1]{fontenc}
\usepackage[latin1]{inputenc}
\usepackage{amsmath}
\usepackage{graphicx}
\usepackage{amssymb}

\makeatletter

\newcommand{\noun}[1]{\textsc{#1}}

\makeatletter

\makeatletter

\makeatletter

\newcommand{\V}{\boldsymbol}

\usepackage{bm}

\usepackage{times}

\usepackage{subfigure}

\usepackage{epsfig}

\usepackage{color}

\definecolor{rouge}{rgb}{1,0,0}

\makeatother

\makeatother

\makeatother

\makeatother

\usepackage{babel}
\makeatother
\begin{document}

\title{Intersublevel Polaron Dephasing in Self-Assembled Quantum Dots\\
 }

\author{E. A. Zibik}

\affiliation{Department of Physics and Astronomy, University of Sheffield, Sheffield
S3 7RH, United Kingdom}

\author{T. Grange}

\affiliation{Laboratoire Pierre Aigrain, Ecole Normale Sup\'{e}rieure, 24 rue
Lhomond, 75231 Paris Cedex 05, France}

\author{B.A. Carpenter}

\affiliation{Department of Physics and Astronomy, University of Sheffield, Sheffield
S3 7RH, United Kingdom}

\author{R. Ferreira}

\author{G. Bastard}

\affiliation{Laboratoire Pierre Aigrain, Ecole Normale Sup\'{e}rieure, 24 rue
Lhomond, 75231 Paris Cedex 05, France}

\author{N.Q. Vinh}

\author{P. J. Phillips}

\affiliation{FOM Institute Rijnhuizen, PO Box 1207, NL-3430 BE, Nieuwegein, The
Netherlands}

\author{M. J. Steer}

\author{M. Hopkinson}

\affiliation{EPSRC National Centre for III-V Technologies, Sheffield S1 3JD, United
Kingdom}

\author{J. W. Cockburn}

\author{M. S. Skolnick}

\author{L. R. Wilson}

\affiliation{Department of Physics and Astronomy, University of Sheffield, Sheffield
S3 7RH, United Kingdom}

\begin{abstract}
Polaron dephasing processes are investigated in InAs/GaAs dots using
far-infrared transient four wave mixing (FWM) spectroscopy. We observe
an oscillatory behaviour in the FWM signal shortly ($<5$ ps) after
resonant excitation of the lowest energy conduction band transition
due to coherent acoustic phonon generation. The subsequent single
exponential decay yields long intraband dephasing times of $~90$
ps. We find excellent agreement between our measured and calculated
FWM dynamics, and show that both real and virtual acoustic phonon
processes are necessary to explain the temperature dependence of the
polarization decay. 
\end{abstract}


\date{\today}

\maketitle
The strong spatial confinement of carriers in semiconductor quantum
dots (QDs) leads to striking differences in the carrier-phonon interaction
compared with systems of higher dimensionality. In particular, the
discrete energy level structure in QDs results in long exciton and
electron dephasing times \cite{Borri01,Birkedal01,Petta05,Greilich06},
making these semiconductor nanostructures highly attractive for implementation
in quantum information processing applications. The study of dephasing
mechanisms in QDs is commonly carried out using transient four wave
mixing (FWM) spectroscopy. Using resonant interband excitation, FWM
measurements have revealed the absorption lineshape of single QDs
to consist of a narrow zero phonon line (ZPL) and an acoustic phonon-related
broadband centred at the same energy. The only intraband FWM study
\cite{Sauvage02-2} involved resonant excitation of high energy transitions
in the valence band of p-doped QDs yielding dephasing times $\sim15$
ps. However it was not possible to determine the dephasing mechanisms
in this case.

Intraband studies of the well-resolved lowest energy conduction band
electron transitions in InAs/GaAs QDs have provided deep insight into
the electron-phonon interaction and carrier relaxation processes in
n-doped samples. Clear evidence of strong coupling between electrons
and phonons, resulting in polaron formation, has been demonstrated
using magneto-transmission measurements \cite{Hameau99}. Ultrafast
studies \cite{Sauvage02,Zibik04} of polaron decay have shown that
the previously assumed 'phonon bottleneck' picture is not valid. Compared
with semiconductor quantum wells, the intraband population relaxation
time in QDs is long ($\sim50$ ps) suggesting relatively long dephasing
times. However there have been no reports of direct dephasing measurements
to date.

In the present letter we present the first investigations of intraband
dephasing in n-doped QDs using degenerate FWM. Our calculations of
the absorption lineshape in this case show marked differences in comparison
with the interband absorption \cite{Besombes01}. The intraband lineshape
consists of peaked acoustic phonon sidebands separated by $\sim1.5$
meV from the ZPL, which corresponds to phonons with wavelength close
to the dot size, and is reminiscent of the lineshape associated with
impurity-bound electron transitions \cite{Toyozawa}. Using pulse
durations short enough to excite both the ZPL and acoustic phonon
sidebands we find damped oscillations in the FWM signal, indicative
of coherent acoustic phonon generation, followed by a single exponential
decay. In contrast with the interband case, where the origin of the
strong temperature dependence of the excitonic linewidth is still
subject to debate, the simple 3-level structure of the lowest energy
conduction band states in InAs QDs permits an accurate simulation
of the temperature dependence of the FWM signal. The excellent agreement
found between experiment and theory, shows that virtual transitions
between the $p$-states is the dominant dephasing mechanism at high
temperature. At low temperature, we have measured an intersublevel
dephasing time of $T_{2}\sim90$ ps. It is also interesting to compare
our results with previous intraband dephasing measurements in higher
dimensional (quantum well) systems \cite{Kaindl01}. Here phonon-mediated
processes are not significant with the intraband dephasing instead
determined by electron-electron interactions, yielding typical dephasing
times $\sim0.3$ ps which are approximately 2 orders of magnitude
faster than for the QD samples studied here. The relatively long intraband
dephasing time in QDs is key to the efficient operation of new types
of mid-infrared QD-based devices, such as intersublevel polaron lasers
\cite{Sauvage06} and may be relevant for potential device applications
such as qubits for quantum information processors \cite{Shervin99}.

The investigated samples were grown on (100) GaAs substrates by molecular
beam epitaxy in the Stranski-Krastranow mode. They comprise 80 layers
of InAs self-assembled QDs separated by 50 nm wide GaAs barriers,
thus preventing both structural and electronic coupling between QD
layers. The polaron transitions were studied between \emph{s}-like
ground (\emph{s}) and \emph{p}-like first excited (\emph{p}) states
within the conduction band. To populate the \emph{s} state, the samples
were delta-doped with Si 2 nm below each QD layer. The doping density
was controlled in such a way that the average doping did not exceed
1 electron per dot (see Ref. \cite{Carpenter06} for more details).
Absorption spectra of the investigated QD samples were studied elsewhere
\cite{Zibik04}. Since our QD samples contain $\sim1$ e/dot only
the ground state is occupied and therefore the incident radiation
polarized along either the {[}0$\bar{1}$1] or {[}011] crystallographic
directions excites a transition from the s state to either the lower
($p_{x}$) or higher ($p_{y}$) energy laterally confined state. The
absorption peaks associated with these transitions are inhomogeneously
broadened by $\sim5$ meV due to the QD size and composition distribution.
The $\Delta_{pp}\sim5$ meV anisotropy splitting between the two peaks
can be explained by QD asymmetry due to piezoelectric field effects
\cite{Stier99,Bester06} and the atomistic symmetry \cite{Bester05}.

\begin{figure}
\begin{centering}
\includegraphics[width=0.4\textwidth,keepaspectratio]{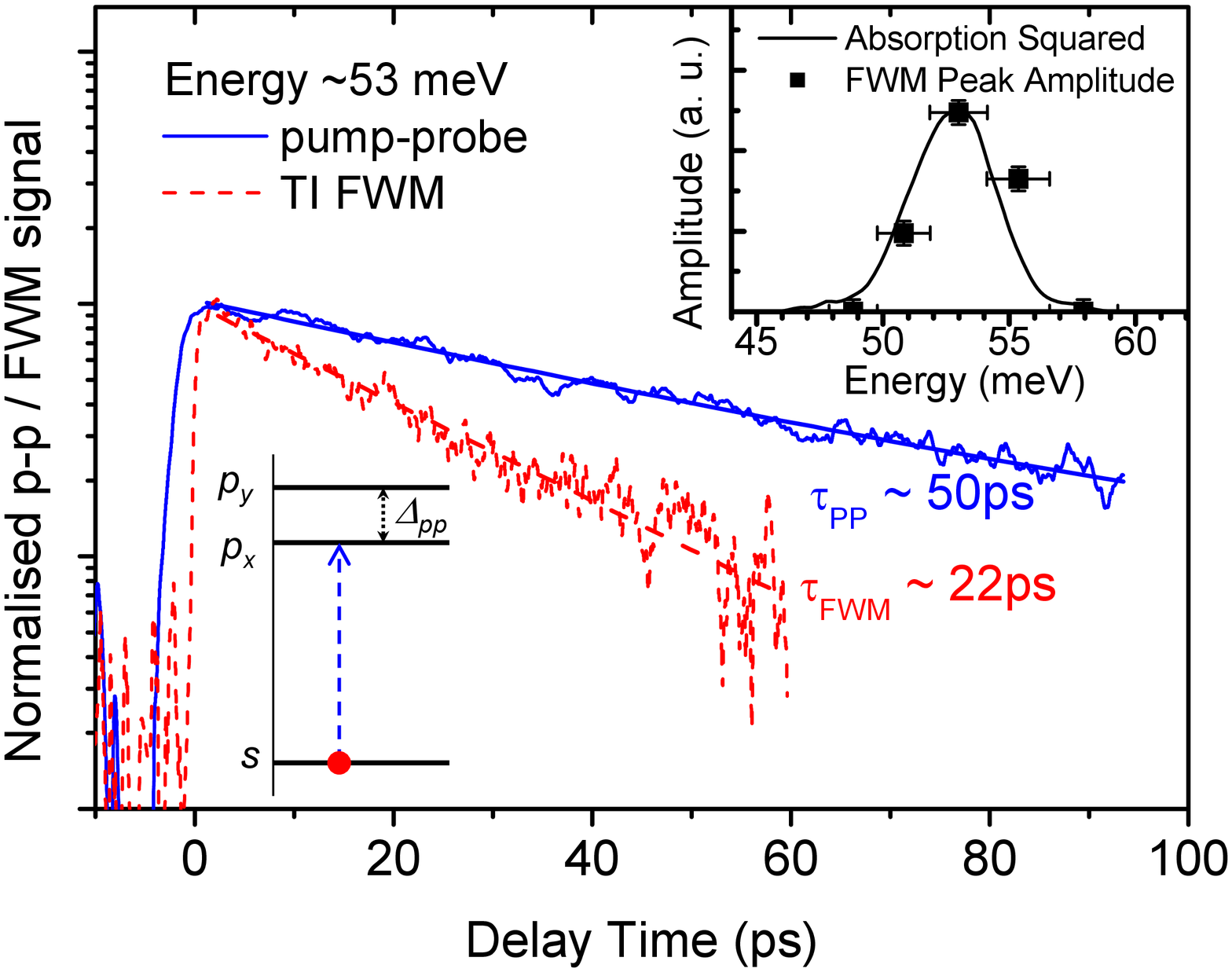} 
\par\end{centering}

\caption{(Color online) Comparison of time integrated FWM (dashed line) and
pump-probe (solid line) signals measured at 10 K and transition energy
of $\sim53$ meV as a function of the delay time between the incoming
pulses. The inset exhibits the absorption squared and FWM signal intensity
at peak maximum vs the transition energy for the same sample. Also
shown is the schematic diagram of the polaron energy structure in
InAs QDs. }

\label{Fig1} 
\end{figure}

We studied the coherent polaron polarizations in QDs using a standard
two-pulse photon echo arrangement in a non-collinear geometry \cite{Zimdars93}.
The far-infrared time-integrated FWM measurements were carried out
using the Dutch free electron laser (FELIX) which provides subpicosecond,
tuneable laser pulses. We used a ratio of 1:2 between the two incoming
pulses with wave vectors $\mathbf{k}_{1}$ and $\mathbf{k}_{2}$,
and the intensity of the third order nonlinear signal was measured
in the $2\mathbf{k}_{2}-\mathbf{k}_{1}$ direction. The applied peak
power density was $\sim50$ W/mm$^{2}$ \cite{Comment1}. The measurements
were performed in the $\chi^{(3)}$ regime, where the intensity of
the FWM signal has a cubic dependence on the excitation intensity.

The comparison between the FWM signal and the pump-probe signal at
the same $s\text{-}p_{x}$ transition energy of $\sim53$ meV is shown
in Fig.~\ref{Fig1}. To verify that the FWM signal arises from excitation
in resonance with the $s\text{-}p_{x}$ transition in the QDs we measured
the spectral dependence of the FWM signal amplitude and find a good
correspondence with the square of the linear $s\text{-}p_{x}$ absorption
signal (inset Fig.~\ref{Fig1}).

Unlike pump-probe, the FWM signal is sensitive not only to changes
in carrier population but also to a decay of coherent optical polarizations,
thus providing a direct measurement of the excited carrier dephasing
time. The decay time of the FWM signal is fitted with a single exponential
curve yielding $\tau_{FWM}\sim22$ ps. In the case of inhomogeneously
broadened transitions the dephasing time is $T_{2}=4\tau_{FWM}$ \cite{Shah},
and thus the low temperature dephasing time of the $s$-$p_{x}$ transition
is $T_{2}\sim88$ ps. This is close to the value $2T_{1}\sim100$
ps deducted from independent pump-probe measurements. The homogenous
linewidth $\Gamma_{2}=2\hbar/T_{2}$ can be decomposed as the sum
of a population relaxation $\Gamma_{1}=\hbar/T_{1}$ and a pure dephasing
$\Gamma_{2}^{*}=2\hbar/T_{2}^{*}$ contributions. The relation $T_{2}\simeq2T_{1}$
indicates that pure dephasing processes are negligable at low temperature.

We find that at low temperature the polaron dephasing time decreases
from $T_{2}\sim88$ ps at an excitation energy $\hbar\omega\sim$53
meV to $T_{2}\sim60$ ps at $\hbar\omega\sim$48 meV \cite{Comment2}.
This is consistent with the energy dependence of the polaron decay
time $T_{1}$ \cite{Li99,Zibik04}, which decreases as the polaron
energy approaches that of the LO-phonon. As we have shown in Ref.~\onlinecite{Zibik04},
the polaron relaxation to the ground state is due to anharmonic disintegration
of polarons into two high energy acoustic phonons, leading to the
following temperature dependence: $\Gamma_{1}=\left[1+N(\hbar\omega_{p}/2)\right]^{2}\hbar/T_{1}^{0}$
, where $N(\hbar\omega)=1/(e^{\hbar\omega/k_{B}T}-1)$ is the Bose
occupation number, $T_{1}^{0}$ is the polaron lifetime at low temperature
and $\hbar\omega_{p}$ is the $s$ to $p_{x}$ energy transition.
But as we shall see below, $\Gamma_{2}\gg\Gamma_{1}$ when the temperature
is increased, indicating that pure dephasing processes become dominant.

\begin{figure}
\begin{centering}
\includegraphics[width=0.4\textwidth]{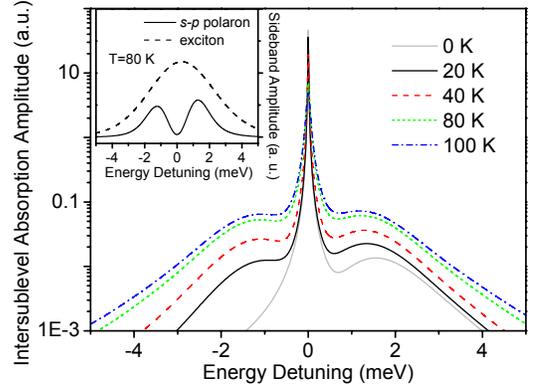} 
\par\end{centering}

\caption{(Color online) Calculated InAs single-dot polaron absorption at different
temperatures. The ZPL transition energy is taken as zero energy. Inset
shows the comparison between the acoustic phonon sidebands for intersublevel
polaron (solid line) and ground state exciton (dashed line) at T=80
K. The ZPL is not considered for clarity.}

\label{Fig2} 
\end{figure}

In order to take into account additional sources of dephasing, we
consider the polaron interaction with bulk-like longitudinal acoustic
(LA) phonon modes: \begin{equation}
V=\sum_{i,j}\sum_{\V{q}}M_{\V{q}}^{ij}(a_{\V{q}}+a_{\V{-q}}^{+})\vert i\rangle\langle j\vert\label{coupling}\end{equation}
 where $M_{\V{q}}^{ij}=D_{c}\sqrt{\frac{\hbar q}{2\rho c_{s}V}}\langle i\vert e^{i\V{q}.\V{r}}\vert j\rangle$
and $\vert i\rangle$ denotes the polaron state with the dominant
component of the electron wavefunction $i=s,p_{x}\text{or }p_{y}$.
The deformation potential constant for the conduction band is taken
as $D_{c}=-7.2~\text{eV}$ \cite{Vurgaftman01}, the sound velocity
is $c_{s}=5000~\text{m.s}^{-1}$ and the density $\rho=5.32~\text{g.cm}^{-3}$.
The diagonal parts within the polaron basis ($i=j$) are treated within
the independent Boson model \cite{mahan}, and results in phonon sidebands
as replicas to the ZPL. The lineshape of the absorption as a function
of the energy detuning $\varepsilon$ is then given by $A(\varepsilon)=Ze^{f}(\varepsilon)$,
where the exponential part is taken in the convolution sense ($e^{f}=\delta+f+f\otimes f/2+...$)
of the function: \begin{equation}
f(\varepsilon)=\sum_{\V{q}}\dfrac{\vert M_{\V{q}}^{p_{x}p_{x}}-M_{\V{q}}^{ss}\vert^{2}}{\varepsilon^{2}}\left[N(|\varepsilon|)+\Theta(\varepsilon)\right]\delta(|\varepsilon|-\hbar\omega_{q})\label{function}\end{equation}
 where $\Theta$ is the Heavyside function. The weight of the ZPL
is given by $Z=exp\left[-\int_{-\infty}^{+\infty}\text{d}\varepsilon f(\varepsilon)\right]$.
Calculation of $A(\varepsilon)$, convolved with a Lorentzian of linewidth
$\Gamma_{2}$ as calculated below, is shown in Fig.~\ref{Fig2} for
different temperatures. Due to cancellations of the contributions
from $s$ and $p$ levels for long wavelength phonons in Eq.~\ref{function},
$f(\varepsilon)$ behaves as $\varepsilon^{4}$ close to zero detuning
$\varepsilon=0$ and is peaked around 1.5 meV, which corresponds to
phonons with wavelength of about the dot size \cite{Comment3}. As
a consequence, instead of a broad peak at the optical transition energy
as in the interband case, this leads to the appearance of two peaks
separated from the ZPL in the absorption spectrum (see inset of Fig.~\ref{Fig2}).

As the ZPL is given by a delta function in Eq.~\ref{function}, the
diagonal parts of the phonon coupling do not contribute to the linewidth
$\Gamma_{2}$. On the other hand, off-diagonal acoustic phonon coupling
between the close-in-energy $p_{x}$ and $p_{y}$ states are expected
to contribute to pure dephasing processes. Taking into account these
off-diagonal interactions with acoustic phonons up to second order
\cite{Muljarov04}, we have calculated analytically the broadening
associated with real and virtual transitions from the $p_{x}$ state
to the $p_{y}$ state: \begin{equation}
\begin{split}\Gamma_{2}^{*}= & \frac{1}{2\pi}\int_{0}^{+\infty}\text{d}\varepsilon\frac{4\Delta_{pp}^{2}}{\left(\varepsilon+\Delta_{pp}\right)^{2}}\\
 & \times\frac{\Gamma_{pp}^{2}(\varepsilon)N(\varepsilon)\left[N(\varepsilon)+1\right]}{\left(\varepsilon-\Delta_{pp}\right)^{2}+\left(\dfrac{\Gamma_{pp}(\varepsilon)\left[N(\varepsilon)+1\right]}{2}\right)^{2}}\end{split}
\label{eqdeph}\end{equation}

where $\Gamma_{pp}(\varepsilon)=2\pi\sum_{\V{q}}\vert M_{\V{q}}^{p_{x}p_{y}}\vert^{2}\delta(\varepsilon-\hbar\omega_{q})$.
The integration of $\Gamma_{2}^{*}$ around $\varepsilon=\Delta_{pp}$
corresponds to the linewidth $N(\Delta_{pp})\Gamma_{pp}(\Delta_{pp})$
of the real transition from $p_{x}$ toward $p_{y}$ by absorption
of acoustic phonon, while phonons with an energy which is not in resonance
with $\Delta_{pp}$ are responsible for virtual transitions, $i.e.$
simultaneous absorption and emission of phonons of same energies but
different wavevectors. By analysing Eq.~\ref{eqdeph} we found that
phonons which contribute mainly to these virtual transitions have
energies $\varepsilon$ between $2$ and $3~\text{meV}$, and their
contribution to the dephasing is proportionnal to $N(\varepsilon)\left[N(\varepsilon)+1\right]$.
Therefore, the full width at half maximum of the single dot homogeneous
line $\Gamma_{2}=\Gamma_{1}+\Gamma_{2}^{*}$ has a strong temperature
dependence.

\begin{figure}
\begin{centering}
\includegraphics[width=0.35\textwidth]{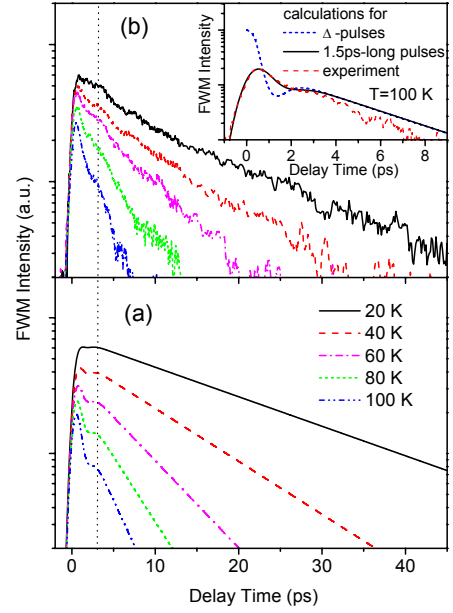} 
\par\end{centering}

\caption{(Color online) Temperature dependent four wave mixing signals: simulations
(a) and experiment (b). Inset: calculated FWM curves at 100 K for
delta pulses (dotted line) and 1.5ps-long pulses compared with experimental
data. }

\label{Fig3} 
\end{figure}

We have calculated the FWM dynamics for excitation in resonance with
the $s$ to $p_{x}$ transition, taking into account polaron decay
to the $s$ state, real and virtual transitions to the $p_{y}$ state,
as well as the presence of phonon sidebands \cite{Vagov03}. The intensity
of the FWM response to a delta pulse as a function of the delay time
between the 2 pulse reads: \begin{equation}
I(t)\propto\Theta(t)\exp\left[-2\Gamma_{2}t-16\int_{-\infty}^{+\infty}\text{d}\varepsilon f(\varepsilon)\sin^{2}\left(\frac{\varepsilon t}{2\hbar}\right)\right]\label{eqfwm}\end{equation}
 Experimental results measured at $\sim53$ meV and calculations of
the FWM are presented in Fig.~\ref{Fig3}. Our calculations reveal
the occurrence of decoherence oscillations between 0 and 5 ps due
to the presence of acoustic phonon sidebands (see Fig.~\ref{Fig3}a),
followed by an exponential decay of the FWM signal on tens of picosecond
time scale ($T_{2}/4$) due to real and virtual polaron transitions.
The period of these oscillations is given approximatively by $h/\varepsilon_{1}$
where $\varepsilon_{1}\sim1.5~\text{meV}$ is the energy separation
between the ZPL and the sideband peaks in Fig.~\ref{Fig2}. These
oscillations, also observed experimentally (Fig.~\ref{Fig3}b), become
more prominent with increasing temperature since the population of
the acoustic phonons increases. In order to make a comparison with
our experimental data, $I(t)$ is convolved by a 1.5 ps-long Gaussian
pulse (see inset of Fig.~\ref{Fig3}b). As the width of the sideband
peaks is also in the $\sim1$ meV energy range, the damping of the
oscillations occur on the same time scale as the oscillation, giving
only one oscillation. Thus, similar to exciton dephasing, the phonon
sidebands are responsible for a fast non-exponential FWM decay on
a picosecond time scale after excitation \cite{Comment4}.

\begin{figure}
\begin{centering}
\includegraphics[width=0.35\textwidth]{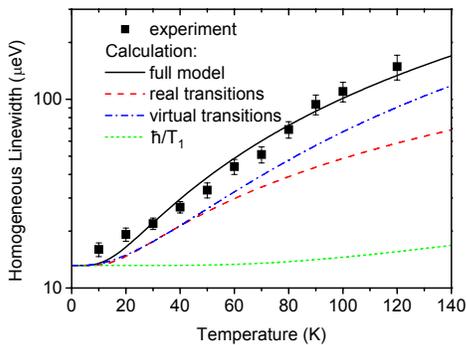} 
\par\end{centering}

\caption{(Color online) Temperature dependence of the polaron linewidth: experiment
(closed squares) and calculation including real (dashed line), virtual
(dash-dotted line) and both virtual and real (solid line) transitions.
Dotted line exhibits the temperature dependence of the population
relaxation.}

\label{Fig4} 
\end{figure}

The decay time of the exponential contribution to the FWM signal reduces
significantly with increasing temperature from $\sim22$ ps at 10
K to $\sim3$ ps at 100 K, resulting in a decrease of the polaron
dephasing time from $\sim88$ ps to $\sim12$ ps over the same temperature
range. In Fig.~\ref{Fig4}, we have plotted the calculated total
linewidth of ZPL $\Gamma_{2}=\Gamma_{1}+\Gamma_{2}^{*}$ (solid line),
which shows very good agreement with the experiment. We also present
the temperature dependence of the polaron linewidth due to virtual
acoustic phonon scattering (dash-dotted line), and the calculated
contribution to the polaron linewidth due to real transitions $N(\Delta_{pp})\Gamma_{pp}(\Delta_{pp})$
(dashed line). Contributions of real and virtual transitions to the
dephasing are comparable up to 60K. However for higher temperatures,
the virtual contribution becomes dominant as it is enhanced quadratically
with temperature. The measured and calculated homogeneous linewidths
$\Gamma_{2}$ increase rapidly from $\sim15$ $\mu$eV up to $\sim150$
$\mu$eV with increasing temperature from 10 to 120 K. This behaviour
contrasts with the weak temperature dependence of the polaron population
relaxation to the ground state $\Gamma_{1}$ over the same temperature
range (dotted line in Fig.~\ref{Fig4}d), indicating that the dephasing
is dominated by pure dephasing for higher temperatures.

In summary, these first FWM studies of intersublevel dephasing in
n-doped quantum dots reveal oscillatory behaviour of the polarization
decay for times $<5$ ps after excitation, followed by a single exponential
decay yielding dephasing times of $~90$ ps at 10 K. The oscillation
at short times arises from coherent acoustic phonon generation due
to resonant excitation of both the zero phonon line and the peaked
acoustic phonon sidebands associated with the intersublevel transition.
We find excellent agreement between our measured and calculated four
wave mixing signals over a wide temperature range, allowing us to
determine the role of real and virtual acoustic phonons in the dephasing
process. Compared with interband studies, intraband investigations
allow a clear picture of dephasing mechanisms in n-doped dots to be
obtained as a result of the simple and well-resolved conduction band
system (with no dark states), providing a new insight into decoherence
mechanisms of electron states in QDs.\\

\acknowledgments Funding was provided by the UK Engineering and Physical
Sciences Research Council (EPSRC) under the EPSRC/FOM agreement, and
grant numbers GR/S76076/01 and GR/T21158/01. The LPA (UMR 8551) is
associated with the CNRS and the Universities Paris 6 and Paris 7.
In addition, we would like to thank Dr B. Redlich and the FELIX staff
for their help and guidance.


\begin{thebibliography}{10}
\bibitem{Comment1} This power is significantly lower compared with
the excitation power of $\sim7$ $kW/mm^{2}$ used in Ref.~\onlinecite{Kaindl01}
for FWM measurements in the case of intersubband transitions in QWs.
\bibitem{Comment2} For this measurement another QD sample with absorption
peak maximum at $\sim48$ meV was studied.
\end{thebibliography}
\end{document}